\documentstyle[12pt,psfig]{article}
\begin{document}
\newcommand{\tphi}{\tilde{\phi}}
\newcommand{\lton}{\stackrel{\large <}{\sim}}
\newcommand{\gton}{\stackrel{\large >}{\sim}}
\newcommand{\beq}{\begin{equation}}
\newcommand{\eeq}[1]{\label{#1} \end{equation}}
\newcommand{\beqar}{\begin{eqnarray}}
\newcommand{\eeqar}[1]{\label{#1} \end{eqnarray}}
\newcommand{\gfm}{{\rm GeV/Fm}^3}
\newcommand{\half}{{\textstyle \frac{1}{2}}}
\newcommand{\vx}{{\bf x}}
\newcommand{\vq}{{\bf q}}
\newcommand{\vp}{{\bf p}}
\newcommand{\vk}{{\bf k}}
\newcommand{\vK}{{\bf K}}
\newcommand{\vv}{{\bf v}}
\newcommand{\vb}{{\bf b}}
\newcommand{\vs}{{\bf s}}
\newcommand{\kn}{ $K_0$ }
\newcommand{\knb}{ $\overline{K}_0$ }
\newcommand{\ks}{ $K_s$ }

\begin{flushright}
CU-TP-832   \\
\end{flushright}
% \vspace*{1.cm}
 \begin{flushleft}
{\Large\bf 
Baryon Number Transport in High Energy Nuclear Collisions
 \footnotetext{ 
*This work was supported by the Director, Office of Energy Research,
Division of Nuclear Physics of the Office of High Energy and
Nuclear Physics of the U.S. Department of Energy under Contract No.
DE-FG02-93ER40764.}}\\[2ex]
{\bf M. Gyulassy$^1$, V. Topor Pop$^{1,2}$, S.E. Vance$^1$}   
\\[2ex]
{\normalsize $^1$Physics Department,
Columbia University,
New York, N.Y. 10027\\
$^2$Institute for Space Sciences,
     P.O.Box MG-6,Bucharest,Romania}\\

%\today
\end{flushleft}
%\vskip 0.9cm
%\begin{quote}  \begin{small}
Abstract: Recent SPS data on the rapidity distribution of protons in
p+S, p+Au and S+S collisions at 200 AGeV and preliminary
Pb+Pb collisions at 160 AGeV 
are compared to HIJING
and VENUS calculations as well as to predictions based on the 
Multi-Chain Model (MCM). The preliminary Pb data suggest
that a linear dependence of the proton rapidity shift
as a function of the nuclear thickness,
as first observed in p+A reactions, may apply 
up to Pb+Pb reactions. 
The observed  rapidity dependence of produced hyperons in both p+A
and A+A reactions however cannot be explained in terms of such  models
without introducing additional non-linear effects.

%[2ex]
% \end{small}   \end{quote}

\section{Introduction}

Preliminary data on  baryon number transport in 
$Pb+Pb\rightarrow p,\bar{p},\Lambda,\bar{\Lambda}$
 \cite{na35_7,na49_4,xu_96,jacobs97,bormann97} at 160 AGeV 
has become available.
These data are of interest  as tests of
 nuclear stopping power
\cite{gyu_95}. 
The first estimates of the baryon stopping power of heavy
nuclei was made by Busza and Goldhaber \cite{bus84} based on 
%$0.2<x_F<0.9$ 
the $A$ dependence of $p+A\rightarrow p+X$ data at fixed $p_\perp=0.3$ GeV.
Baryon stopping power refers to the transport of baryon number away
from the nuclear fragmentation regions and is measured in terms of the
 single inclusive rapidity distribution of protons
and hyperons. In ref. \cite{date},
the estimates were refined by taking into account exact Glauber
geometry and applying  the Multi-Chain Model (MCM) parameterization of
baryon number transport.  For a review of the average rapidity loss
for $p+p$, and $A+A$ collisions at beam momenta 11.6, 14.6 and 200
GeV/c per nucleon see  ref.\cite{vide_95}.  Until recently,
 the most  complete information on  nuclear stopping power
was limited to the $p+Ag$  data of Toothacker et al. \cite{toot_87}.

Systematic data from heavy-ion collisions
\cite{na35_7,na49_4,xu_96,jacobs97,bormann97,na35_1,na49_1,harris} provide
new information about the  nuclear stopping power.
  Important
information about baryon inelasticity also comes from the analysis of
veto and
transverse energies distributions \cite{na35_3}-\cite{wa80_4}.

As summarized in \cite{gyu_95}, 
baryon number transport is one of the key observable
that has been debated for some time in connection
with the  ongoing search for nonlinear dynamical phenomena in nuclear
reactions. One source of nonlinear behavior may arise 
if a quark-gluon plasma is formed in such reactions.
In this connection, the observed strangeness enhancement and hyperon 
production data
\cite{na35_7,bormann97,na35_1,na35_3},\cite{na35_2}-\cite{na35_6}
 have stimulated particular interest\cite{raf1,gazd_96}.
In \cite{top_95} it was pointed out, however, that the unusual difference
between hyperon production in $p+p$ and minimum bias $p+S$ 
may reflect more the onset of novel {\em non-equilibrium} 
dynamical mechanisms.

The breakdown of the linear dependence of baryon rapidity shifts as a
function of the mean collision number, $\nu\sim A^{1/3}$, is  one
of the obvious  places to look for new phenomena. The linearity
assumption underlying MCM and many Monte Carlo event generators is
 motivated by the phenomenology  of string formation and
fragmentation\cite{fritiof,dpm94,hijing,venus}. Such linear behavior is also
expected if the radiative energy loss per unit length is a constant in
nuclei\cite{gyuwang}. 

 Several recent theoretical developments
have raised, however,  new possibilities
for  nonlinear dynamics
associated with baryon number transport.
In one development \cite{dokshitz}, a
non-linear energy loss of gluons in nuclei
was predicted as a function of the nuclear
thickness, $L$
\beq
{\Delta E_g}\approx \frac{1}{8} \alpha_s N_c L\; \langle p_\perp^2\rangle_L
\sim 15\; {\rm GeV}\; \left( \frac{L}{10\;{\rm fm}} \right)^2
\; \; .\eeq{dok}
This energy loss could, under appropriate conditions,
lead to enhanced rapidity shifts
of hadrons passing through nuclei. 

A second possible source of nonlinearity was suggested in
ref.\cite{khar96} based on the Rossi-Veneziano\cite{rossi77}
 baryon-junction Regge
exchange model. This mechanism can give rise to large rapidity shifts
of the baryon number offset by  only a modest enhancement of 
the forward final pion rapidity density.
This mechanism  could even modify significantly baryon
transport up to RHIC energies
because the assumed junction Regge intercept, $\alpha_J(0)=0.5$,
leads  to  a slow suppression of this mechanism 
with increasing energy ($\propto s^{-1/4}$).

An even more exotic baryon transport mechanism was proposed in
\cite{kopel} based on a  variant\cite{nussinov} of
the above baryon-junction exchange model.  
In that variant the junction trajectory is assumed to have
unit intercept and thus leads to an energy independent uniform rapidity
density component of the inclusive proton yield.

Given the new data and the above theoretical speculations,
it is appropriate to take analyze carefully the available information
on baryon transport 
and hyperon production. We  therefore 
take into account the older $p,\bar{p}+Ag$
data at 100A GeV \cite{toot_87}, the new $p+A$ \cite{na35_5},
\cite{na35_7} at 200 GeV, the $S+A$ at 200
 AGeV\cite{na35_7,na35_4,na35_5}, and the $Pb+Pb$ data at 160A GeV 
 \cite{na49_4,jacobs97,bormann97,na49_1,na35_6}.
 In a previous paper, we concentrated on the
anomalous hyperon production processes\cite{top_95}.
 Here we concentrate on the proton rapidity shifts in nuclear
collisions. We test whether  nuclear stopping power extrapolates linearly
from $p+A$ to $Pb+Pb$.
In this analysis we recall  the
predictions of the Multi-Chain Model\cite{date} (MCM)
as well as  utilize the Monte
Carlo HIJING1.3 \cite{hijing} and VENUS4.12 \cite{venus} models as in 
\cite{top_95}.

\section{The Multi Chain Model}

Since the HIJING and VENUS models were discussed extensively
in \cite{top_95}, we recall here only the essential  elements of the MCM
to be used in the present  analysis.

The single inclusive proton rapidity distribution in $B+A\rightarrow p+X$
reactions is given in general  by Glauber theory 
as
\beq
\frac{dN^{BA\rightarrow pX}}{dy}
= \sum_{m=1}^B \sum_{n=1}^A P_{BA}(m,n) Q_{m,n}(y) \; \; , 
\eeq{5.2}
where 
the tedious but well understood nuclear
geometry is separated from the
multiple collision dynamics encoded in the functions, $Q_{m,n}(y)$.
The probability that a group of $m$
 projectile nucleons multiple scatter with  $n$ target
nucleons is given by
\beq
P_{BA}(m,n)=\int \frac{d^2\vb}{\sigma_{AB}} {\cal B}(\vb)
\int\frac{d^2\vs}{\sigma_{in}}
P_B(m,N_B(\vs)/B)P_A(n,N_A(\vb-\vs)/A)
\; \; , \eeq{pab}
where $P_A(m,x)=C_{m,A}x^m(1-x)^{A-m}$ is the binomial distribution.
The mean number of inelastic collision in 
nucleus $A$ at impact parameter $\vb$
in terms of the diffuse nuclear density and inelastic $pp$ cross section,
$\sigma_{in}\approx 32$ mb by
$ N_A(\vb)=\sigma_{in}\int dz\;\rho_A(z,\vb)$. 
The impact parameter profile function ${\cal B}$ is included above
to account for the experimental trigger bias (e.g. veto or transverse
energy cuts). For minimum bias events, ${\cal B}=1$.
For a given profile, the total reaction cross section
is
\beq
\sigma_{AB}=\int d^2\vb {\cal B}(\vb)\left(1-(1-\int d^2\vs
N_B(\vs)N_A(\vb-\vs)/(AB\sigma_{in}))^{BA} \right)
\; \; . \eeq{sigtot}

The main simplifying assumption of the MCM is that 
of independent fragmentation:
\beq
Q_{m,n}(y)\approx m Q_n(Y-y) +n Q_m(y)
\; \; ,
 \eeq{ind}
where $Y$ is the rapidity difference between the projectile and target.
As emphasized in \cite{date}, this linear superposition ansatz is far from
obvious, and one of the important questions awaiting the recent heavy
ion data is whether this breaks down for
sufficiently heavy nuclear collisions\cite{gyu_95}.
With this ansatz the rapidity distribution simplifies 
to \beq
\frac{dN^{BA\rightarrow pX}}{dy}
= r_B W_B\sum_{n=1}^A P_{B/A}(n) Q_n(Y-y) + r_A W_A
\sum_{m=1}^B P_{A/B}(m) Q_n(m)
\; \; , \eeq{gl}
where $r_B=(Z_B f +N_B(1-f))/B$ 
is the fraction of projectile baryons that fragment
into protons, and where $W_B$ is the average number of 
wounded baryons in nucleus $B$:,
\beq
W_B=\int \frac{d^2\vb}{\sigma_{AB}}
 {\cal B}(\vb)\int\frac{d^2\vs}{\sigma_{in}}
N_B(\vs)\{1-[1-N_A(\vb-\vs)/A]^A\}
\eeq{w}
and $P_{B/A}(n)$ is the fraction of them that
interact with $n$ target nucleons.
 Typically\cite{date}, the fraction of incident protons that
remain protons
is $f\approx 0.53$ away from the diffractive
peaks.

What remains then to be determined are the dynamical fragmentation functions,
 $Q_n(y)$, specifying the rapidity distribution of target baryons that have
 suffered $n$ inelastic interactions. 
For $n=1$, i.e., $p+p$, the MCM assumes  the simple form
\beq
Q_1(y,\alpha) = e^{-y}
\; \; , \eeq{q1}
as  consistent with the flat $dN/dx$ distribution  observed
away from the 
diffractive peak ($x\sim 1$). For $n\ge 2$
the multiple collision contribution to baryon rapidity
transport is parameterized in the MCM model by
a one parameter class of  functions:
\beq
Q_n(y,\alpha)=\left(\frac{\alpha}{\alpha-1}\right)^{n-1}\left[
e^{-y}-e^{-\alpha y}\sum_{m=0}^{n-2} \frac{1}{m!}(\alpha-1)^m y^{m} \right]
\; \; . \eeq{qn}
These functions arise using  a scaling algorithm\cite{date} in which
the probability density that spectator partons retain a fraction $z$
of the total light-cone momentum after an inelastic interaction
is $\alpha z^{\alpha-1}$. This ansatz implies a geometric
scaling of the spectator energy fraction
moments $\langle z^p\rangle_{Fn} = (\alpha/(\alpha+p))^n$
and a linear scaling of the mean rapidity shift
with collision number $n$:
\beq
\langle y\rangle_n =1 + (n-1)/\alpha
\; \; .\eeq{dely}
Detailed fits to the Barton et al\cite{bart83}  $p+A\rightarrow p+X$ 
data at $100$ GeV,  gave 
$\alpha=3\pm 1$. Other models\cite{bus84} can also
achieve good fits to the existing  $p+A$ data 
with different parameterizations.
For example, a recent fit with the constituent quark model
was presented  in \cite{takagi}.
The advantage of the MCM approach is in its simplicity, 
reducing the problem of  the nuclear stopping power to one phenomenological
parameter, $\alpha$. Recall that the naive incoherent
cascade limit corresponds to $\alpha=1$, leading to one unit rapidity shift
per interaction. The empirical $\alpha=3$ arises because the inelasticity
 per interaction 
inside nuclear targets is reduced by the finite
formation time of secondary fragments.

We note that the scaling approximation used above certainly breaks down
for lower energies where the rapidity gap, $Y$, between projectile and target
becomes $\lton 3$. In the 
following we simply cutoff $Q_n$ with a $\theta(Y-y)$
factor and normalize $Q_n$ in that finite interval. In \cite{date}
a slightly different finite energy cutoff prescription was used.
Fortunately, for $E_{lab}>100$ GeV such cutoff effects are unimportant.
 
At the $pp$ level we can  test for the possibility
of novel baryon junction transport\cite{khar96}.
If the probability of junction exchange is $\epsilon_J$ and its
Regge intercept is $\alpha_J$, then $Q_1$ would be modified from (\ref{q1})
to 
\beq
Q_1(y)= (1-\epsilon_J) e^{-y} + \epsilon_J (1-\alpha_J) e^{-(1-\alpha_J)y} 
\;\; . \eeq{q1khar}
Actually, the contribution of junction exchange close to the fragmentation
regions is not well determined and the above form may only apply for
$y\gg 1$. Given the above form,
the final $p-\bar{p}$ rapidity distribution in the cm
with a rapidity
gap $2Y$ would be
\beq
\frac{dN^{pp\rightarrow pX}}{dy}= r_p\left((1-\epsilon_J)\
\frac{\cosh(y)}{\sinh(Y)}
+\epsilon_J (1-\alpha_J) \frac{\cosh((1-\alpha_J)y)}{\sinh((1-\alpha_J)Y)} 
\right)
\; \; . \eeq{khar}
%Fig1

In figure 1, the  valence
proton, $p-\bar{p}$, distribution from $pp$ collisions at 400 GeV\cite{pp400}
is shown compared to MCM and HIJING. These same data
were  used  to test VENUS in \cite{venus}. The mid-rapidity yield
is underestimated by HIJING by a factor of two, while
the VENUS proton fragmentation scheme fits  better\cite{venus}. 
Both VENUS and HIJING
over-predict the forward $y\sim6$ yields. The excellent
agreement with the simple  MCM form, $\cosh(y-Y)$,
with $r_p=0.53$  leaves little  room 
for exotic baryon exchange
contributions at this energy. 
At higher energies, $\surd s =53 AGeV$, Ref.\cite{khar96} finds evidence
for taking $\epsilon_J>0$. However, Fig. 1 demonstrated 
 conventional baryon
trajectory exchange reproduces accurately the observed valence proton rapidity
distribution at this lower energy. 
This point is important  since in \cite{khar96} it was hinted
that at SPS energies such conventional mechanism 
at least as expressed in the  Dual Parton Model
\cite{dpm94} could not reproduce the experimental stopping power
in nucleus-nucleus reactions. Part of the problem in DPM and HIJING
discrepancy with $pp\rightarrow pX$ data can  be traced to the
assumed diquark fragmentation schemes. We return to
this point in section 5.

%One of our main conclusions below,
%however is that  the MCM model using (\ref{q1}) and 
%(\ref{qn}) with parameters fixed 
%by $pp$ and $pA$ data do in fact account quantitatively
%for the observed nuclear stopping up to Pb+Pb.

\section{Baryon transport in $p+A$}

In Figure 2. we compare calculations with the 100 GeV
$p,\bar{p}+Ag\rightarrow pX$ data\cite{toot_87}. 
Part (a) shows the  $dN/dy$ distribution
in  the projectile fragmentation domain ($\Delta y<2)$
and the higher but narrower $dN/dy$
distribution in the target fragmentation domain ($\Delta y>4$).
 Both the MCM prediction
(with $\alpha=3$) and the VENUS model are seen
 to reproduce the leading proton distribution
within the  experimental errors. 
HIJING leads to a similar mean rapidity shift but
is  distributed more narrowly  about the mean. This is due to the default
diquark fragmentation scheme in JETSET\cite{fritiof} which is 
 used in HIJING. 
 VENUS does not use the JETSET scheme and has been adjusted 
to reproduce the flat
 fragmentation region\cite{venus}.

The target region is isolated  in part (b) through
the $\bar{p}+A\rightarrow p+X$ channel. 
Since  MCM and HIJING only account for the wounded baryons,
the narrow nuclear enhancement around the target rapidity ($\Delta y=5.3$)
can only be reproduced by VENUS, which incorporates final state interactions.
However, the main interest is  the probability
that a target proton emerges near the projectile rapidity
$\Delta y\sim 1$ in Fig.2b. While the data fluctuate
greatly in that region, they are consistent with the expected 
exponential fall from eq.(\ref{q1}). As in Figure 1, there is little need here
 to invoke  a junction exchange contribution\cite{khar96}
with intercept $1/2$ that would lead to a $\exp[y/2]$ rather
than the conventional $\exp[y]$ tail in this opposite fragmentation region.

In Fig. 2c the projectile fragmentation region is
isolated through the $\bar{p}+A\rightarrow\bar{p}+X$ channel.
The MCM model is consistent with the data while both HIJING and VENUS tend
to overestimate this distribution just as both over-predicted
the $y\sim 6$ region. 
In Fig. 2d 
the $A$ dependence of the MCM baryon distribution
is illustrated.
We recall that in MCM, $dN/dy(y\rightarrow 0)=r_p P_A(1)$ is completely fixed
by the geometrical probability that the projectile proton interacts
only once and the probability, $r_p$ that the proton 
remains a proton after
fragmentation. (As in Fig.1 ,  $r_p=0.53$ \cite{date} here.)
%Figure 2

In Figure 3, we compare the models to preliminary $p+A$ data 
reported from NA35\cite{na35_7}.
Parts (a) and (c) show the $p-\bar{p}$ distributions for $S$ and $Au$ targets
respectively. The peculiar feature in (a) relative to previous data is 
%Figure 3 
that in this case HIJING best reproduces  the projectile
fragmentation peak in $p+S$. In this case  the MCM 
distribution is too flat. The mean rapidity shift in all three models is about
the same ($\Delta y\approx 1.3$) in accord with previous systematics
\cite{bus84}, but the narrow peak at $y=5$ is unexpected for minimum bias
events given in Figures 1 and 2. 
The new $p+Au$ minimum bias data in part (d) also indicates a greater
stopping power than encoded in the models
HIJING and VENUS with a significantly suppressed yield beyond
$y>4$ and an enhanced yield below $y<4$. The calculated minimum bias rapidity
shift in MCM is $\Delta y\approx 2$ while the data indicate perhaps
$\Delta y\approx 2.5$. Such a large rapidity shift
is achieved only in the
 extrapolated most central $p+A$ reactions in the older 
data\cite{bart83}.

There appears therefore to be a discrepancy
between the data in Fig.3a and the older data\cite{bart83}.
 Unfortunately the data in Fig.2
have too low  statistical significance  to settle this problem.
Note that in these plots the unobserved target fragmentation regions
are artificially cutoff below $y<0.5$.
Upcoming $p+A$ data with NA49 will hopefully
clarify this ambiguous situation.
 
 As seen in Fig. 3, the net  
$\Lambda-\bar{\Lambda}$ hyperon
distributions is even in greater disagreement
  with respect to both VENUS and HIJING calculations.
In ref.\cite{top_95} the absence of $\Lambda$ fragments beyond
$y>5$ in $p+S$ and beyond $y>4$ in $p+Au$ was emphasized
to be anomalous relative to $pp\rightarrow\Lambda$ data
and $\mu p\rightarrow \Lambda$ data previously analyzed in \cite{top_95,venus}.
In those $p$ target reactions,
the $\Lambda$ rapidity  distribution closely mirrors the proton fragment
distribution with reduced normalization
due to the expected suppression of  strange flavor production.
The suppression of the forward $\Lambda$ relative to forward $p$ 
 cannot be accounted for even with the double string mechanism
added into VENUS. The lack of any measured $\Lambda$'s
beyond $y>4$ in minimum bias $p+Au$ is especially peculiar.
We thus re-confirm that hyperon transport in $p+A$ is
anomalous relative to non-strange baryons, and is 
already evident in minimum bias
$p+S$ reactions, where on the average,
the incident protons only interacts 
with two target nucleons. 

\section{Baryon Number Flow in $A+A$}

We turn next to the distribution of valence baryons
in $A+A$ collisions at SPS for which some preliminary data have
 become available from NA49\cite{na49_4,jacobs97,na49_1} and
NA44\cite{xu_96}. As noted in \cite{gyu_95} data on 
nuclear stopping power in $Pb+Pb$ collisions is long awaited
as a critical test of nuclear transport models
and to  constrain
 the maximal baryon densities achieved in such reactions.
%Figure 4 
 
Figure 4 compares the spectrum of pion and participants protons
in $S+S$ at 200 AGeV  \cite{na35_4} and $Pb+Pb$ reactions
at $160$ AGeV \cite{na49_4,xu_96,na49_1}.
The various data sets from NA35 for $S+S\rightarrow \pi^- + X$
correspond to different centrality triggers \cite{na35_4},
with the higher one corresponding to a more severe veto trigger cut
 (see reference \cite{na35_4} for more details).
We see that the negative pion rapidity densities
are well accounted for by both 
 HIJING \cite{hijing} and VENUS \cite{venus}.
This is largely due to the fact that the pion distribution
simply grows linearly with the atomic number between $S+S$ and $Pb+Pb$.
The centrality trigger was implemented in the above calculations via an 
impact parameter cut of $b<1$ fm. For the MCM calculation the
impact parameter cut for $Pb+Pb$ was taken to be 3.3 fm 
 to simulate more closely the 5-6\% centrality
trigger\cite{na49_4}.

The significant central rapidity minimum  in the $p-\bar{p}$  $dN/dy$
 predicted by HIJING
is traced back here to the excess central minimum   in $pp$ in Fig.1
characteristic of FRITIOF type string models. 
By adjusting the fragmentation functions
that fit the $pp$ central rapidity
region better, VENUS, can avoid the suppression
of the mid-rapidity protons that  HIJING and other  models
using JETSET predict. On the other hand, MCM, which
 fits well both the $pp$ and the
$pA$ data as shown in Figs. 1,2, is seen
to reproduce well the central $p-\bar{p}$
distribution  in both $SS$ and $PbPb$.

Figures 4a,c demonstrate the insensitivity
of the pion distribution to the underlying baryon number flow.
In particular, the rather large difference between
the proton distribution in HIJING and VENUS in Fig.4d is
contrasted by the much more modest difference of the pion
distributions in Fig.4c. As we show in section 5,
 the forward energy flux is a more sensitive measure of the 
 energy degradation difference
between the models.

Our main conclusion is that the central region 
baryon number transport in $A+A$
can be well understood on the basis of a linear extrapolation
of the mean rapidity
shift as a function of the collision number 
as given by eq.(\ref{gl}). The main difference between the valence
distribution functions in $pA$ and $BA$ arises simply from
the variation of the Glauber geometrical probabilities of multiple
interactions in the different projectile and target combinations.

An important caveat to the above conclusion is the
somewhat narrower rapidity distribution of the most recent
analysis\cite{jacobs97} on $Pb+Pb\rightarrow (p-\bar{p})$ than
reported in \cite{na49_4}. This difference is shown by comparing the solid dots
to the solid triangles in Fig.4d.If in the final analysis, the \cite{jacobs97}
narrow distribution is confirmed, then the $Pb+Pb$ baryon stopping power
would have to be regarded as anomalous relative to 
$p+A$\cite{bus84,toot_87}has well since then an additional 1/2 unit rapidity
shift would be required over that predicted by MCM nor VENUS.
In \cite{jacobs97}, it was estimated that the net integrated proton number
implied by the recent analysis is close to $2\times Z(Pb)$. This itself is
peculiar since the one expects more protons from $n\rightarrow p$ exchange.
At this time the central rapidity region is most reliable since
another experiment \cite{xu_96} and alternate TOF measurements converge
within errors in that region and are closest to the theoretical expectations
based on $p+A$ systematics. Nevertheless, it will be important to follow up the
high rapidity tails assess the significance of the present preliminary
indications.

Figure 5 shows that modulo the extreme fragmentation regions,
the enhanced hypron yield in $SS$ and $SAu$
can be understood in terms of the VENUS model double string hypothesis
and in strong disagreement with extrapolated yields using HIJING from $pp$
reactions.  The factor of two enhancement of hyperon
production is already needed in minimum bias $pS$ in Fig. 3b,
but is even more evident in Fig.5. The excess rapidity
shift of the hyperons relative to non-strange baryons is also clear from 
Fig.5 since the peaks are shifted approximately one unit
of rapidity further than those  calculated  for protons in Fig.4b.
This enhanced hyperon transport is currently not explained
by any of these models. In the preliminary $Pb+Pb$ analysis\cite{bormann97}
a truly astonishing sharply peaked $\Lambda$ distribution was suggested
with $dN(\lambda)/dy=23\pm3$. If confirmed,  $\Lambda/p\approx 0.7$ would be
one of the most remarkable signature of novel phenomena in the heavy ion
reaction.

Since the MCM can account for at least the central region, 
non-strange baryon flow up to
SPS energies, it is of interest to extend the calculation
up to RHIC energies, where the rapidity gap opens up to over 10 units.
In Figure 6 we show the expected valence proton distribution
expected at RHIC compared to present energies for $Pb+Pb$ at $b<1$ fm
central collisions.
From this we estimate that the central baryon density
that can be conservatively expected at RHIC is 
\beq
\rho_B\approx \frac{3}{2 } \frac{dN^{p-\bar{p}}}{dy}\frac{1}{r_p \pi R^2\tau}
\approx \rho_0\; (0.7\;{fm}/\tau)
\; \; ,\eeq{rhoB}
  where $\rho_0\approx0.17$ fm$^{-3}$.
   
\section{Energy Degradation versus Baryon Number Transport}

A  question related to baryon transport is the rapidity distribution
of energy lost by the valence baryons.
Do a few high rapidity pions carry
away  the missing baryon energy or is the energy shared between
a large number of slower pions? Data on the forward veto calorimeter
distribution shed  useful light on this problem
as was first emphasized in the analysis of  WA80 data\cite{wa80_2}.

In Figure 7 the veto calorimeter cross section is shown
for   NA35 $S+S$ \cite{na35_3} (Fig. 7b) and  
    NA49 $Pb+Pb$ \cite{na49_2} (Fig. 7d ).
The veto calorimeter measures energy in a narrow angular cone with
$\Delta\theta\sim 0.3^o$.
%Figure 7
   In both cases the well known horseback shape
follows directly from Glauber  geometry of spectator nucleons
and is reproduced by VENUS and HIJING.
However, the tail region at small $E_{VETO}$ is sensitive to the
energy degradation in central collisions.
In Fig. 7 a,c the dependence of the contribution
from collisions in the range $b<1$ fm is illustrated
in both VENUS and HIJING models. For central collisions,
 VENUS,in
closer agreement with the data, 
 displays greater energy degradation than HIJING.

The correlation of the mean veto energy 
with the total multiplicity, participant nucleon number and
impact parameter in both models is  shown in Fig. 8.
%Figure 8
Fig. 8d demonstrated the equivalence of the
Glauber multiple-collision geometry used in both models.
However, the other parts show
that the veto energy is systematically higher (and hence the energy degradation lower)in the HIJING model.
Figure 9 shows the analogous correlations for the Pb+Pb reaction.
%Figure 9

While at first sight it may seem obvious that greater baryon stopping implies
greater energy degradation, we show in Figure 10 that such a correlation
cannot be  taken  {\em a priori} for granted
given the uncertainty in the soft hadronization  mechanism.
In an older version of the diquark fragmentation
scheme used in the version JETSET6.3\cite{jetset63}
it was possible modify the baryon fragmentation through the
parameter $p=PAR(52)$ in the LUDAT1 common block. 
This parameter controls the  momentum fraction of the junction $J$ quark
of a given diquark through
\beq
f(x_J) \propto x_J(1-x_J)^p\; \; ,\; \; \langle x_J\rangle=\frac{2}{3+p}
\eeq{par52}
Increase of the parameter $p$ beyond the default value $p=1$
obviously softens  the diquark fragmentation function and leads to larger
baryon rapidity shifts.
Using the ATTILA\cite{atti} version of the
FRITIOF model\cite{fritiof} it is possible  to explore
the consequences of varying this nonperturbative  model parameter.
The results are shown in Fig.10. We note that HIJING1.3\cite{hijing}
utilizes the new particle conventions of  JETSET7.2\cite{pythia}, 
which are incompatible
with the previous version 6.3.
Unfortunately, in  version 7.2 and higher, it is
no longer possible  to vary par(52). The once option 
(ihpr2(11)=2 in HIJING) to switch to the JETSET ``popcorn''
fragmentation scheme unfortunately leads to even greater discrepancy
with respect to $pp\rightarrow pX$ data in Fig.1. 

It is seen from Fig.10 that setting $p=10$ for $Pb+Pb$ reactions leads to a
very large modification of the net proton distribution, from the default
minimum to a maximum.  In this case $\langle x_J\rangle\approx 0.15$.  In
comparison, the MCM model leads to $\langle x\rangle=
(\alpha/(1+\alpha))^{n-1}/2\sim (3/4)^3/2 \approx 0.2$ for central $Pb+Pb$
where the {\em average} number of collision per wounded nucleon is $\langle \nu
\rangle \approx 4.2$.  The MCM curve from Fig.4 is also shown for comparison.

 We find, therefore, that softening the diquark fragmentation can lead to
 similar baryon number transport as MCM and VENUS.  Recall that VENUS
 incorporates also a diquark breakup component in its fragmentation scheme.
 The important point here is not that a variation of $p$ can simulate more
 closely the baryon transport in VENUS and MCM, but that this model shows 
 a priori no correlation between the energy degradation and baryon number
 transport can be assumed.  In Fig. 10b the veto calorimeter cross section is
 essentially independent of $p$ parameter.  In this model the missing baryon
 energy is carried away into the veto calorimeter by a few high rapidity pions
 that fragment from the leading quark. The mechanism in VENUS that leads to
 greater energy degradation with increasing baryon rapidity shift is the
 dynamical assumption of the occurrence of a double rather than a single string
 in a fraction of the events that lead to large baryon rapidity shifts.
From this analysis we conclude that the diquark mechanism of JETSET
must be replaced in HIJING to enable the simultaneous reproduction of both the
rapidity distributions and the veto calorimeter systematics.

%Figure 10

\section{Conclusions}

One of our  conclusions is that the nuclear stopping power, extracted
from  $p+A$ data\cite{bus84} and extrapolated linearly via the MCM model to
nuclear collisions, accounts 
 quantitatively  for the observed mid-rapidity baryon number
transport up to central $Pb+Pb$ reactions at 160 AGeV  
AGeV\cite{na35_7,na49_4,xu_96}.  The main difference between the valence baryon
distribution functions in $pA$ and $BA$ arises from the variation of the
Glauber geometrical probabilities of multiple interactions.  The pion
rapidity
distributions are insensitive to the details of the energy degradation
associated with the baryon number transport
and are proportional to the wounded nucleon number.  

The forward veto calorimeter data is a sensitive probe
of energy degradation and amplifies subtle variations of the angular
distribution of fast pions.  
The HIJING model and its variants (with modified diquark fragmentation
via eq.(\ref{par52}), cannot account for the veto cross sections.
 VENUS, on the other hand, reproduces both the 
baryon distribution and veto cross
sections well. It remains to find out whether 
the novel double string mechanism
assumed in VENUS is critical to both observations,
or could the veto cross sections simply reflect an enhanced final state
interaction effect in the spectator regions.
In any
case, the strongly non-linear dependence of hyperon production definitely
requires new dynamical mechanism that becomes operative already in $p+S$
reactions. The VENUS model 
identifies this new mechanism with the double string formation.

However, the 
large hyperon rapidity shift in the preliminary $p+S$ and $S+S$ data
remain unaccounted for even with the present double string mechanism
in VENUS 4.12.
Not only is there a large enhancement of midrapidity 
hyperon production in $p+S$ and $S+S$, but there is a rapidity mismatch
between strange and non-strange baryons.
Only the non-strange baryon transport is linear with $A^{1/3}$
within the present errors of the experiments. The provocative
preliminary data of \cite{jacobs97,bormann97} also point in the same direction.

Hyperon production in $Pb+Pb$ is however not entirely settled experimentally.
The NA44 data in Fig. 4d include a substantial fraction of the net hyperon
production while in NA49, the hyperon contribution has been subtracted out. The
present experimental and systematic errors are too large to rule out a
midrapidity maximum of the net baryon distribution.  Our conclusion on
linearity of baryon transport is thus subject to the above caveats.
The present analysis in any case demonstrates  the necessity of modifying the
baryon fragmentation part of HIJING, FRITIOF, and DPM models.

\section{Acknowledgments}
One of us (VTP) thanks M. Morando
 and R. A. Ricci for kind hospitality at INFN - Sezione di 
 Padova, Italy , where part of the calculations were
 performed.  Discussions with J.Harris, P. Jacobs,  D. Kharzeev,
B. Kopeliovich, S. Margetis, W. Willis,  and N. Xu are gratefully acknowledged.
We also thank K. Werner for permission to use the VENUS code.

\newpage
%Fig1
\begin{figure}[htb]
%\vspace{3cm}
\hspace{0.6in}
\psfig{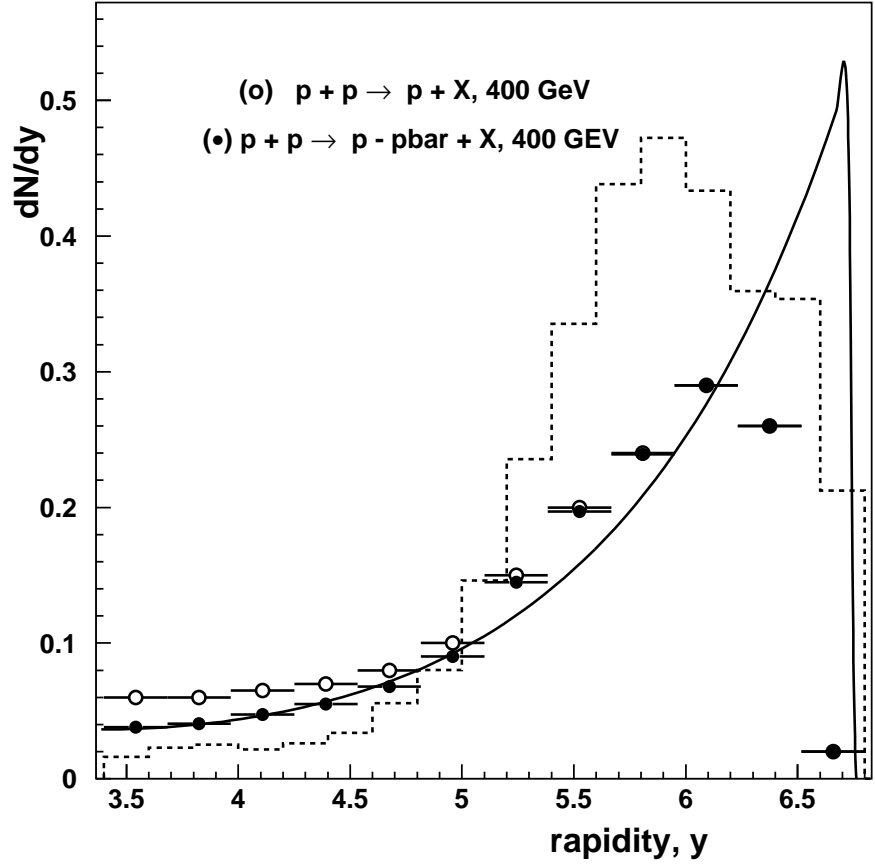}
\caption{Solid circles show the valence proton  rapidity distribution ($dn_p/dy-dn_{\bar{p}}/dy$)
in non-diffractive $pp$ reactions at 400 GeV \protect{\cite{pp400}},
and open circles show the net proton distribution.
The solid curve  corresponds to the default Multi
Chain model\protect{\cite{date}} fragmentation function eq. 
(\protect{\ref{q1}}) with
$r_p=0.53$. The dashed histogram is obtained using the HIJING1.3
\protect{\cite{hijing}} code.}
\end{figure}
\newpage
%Figure 2
\begin{figure}[h]
\vspace{1cm}
\hspace{0.5in}
\psfig{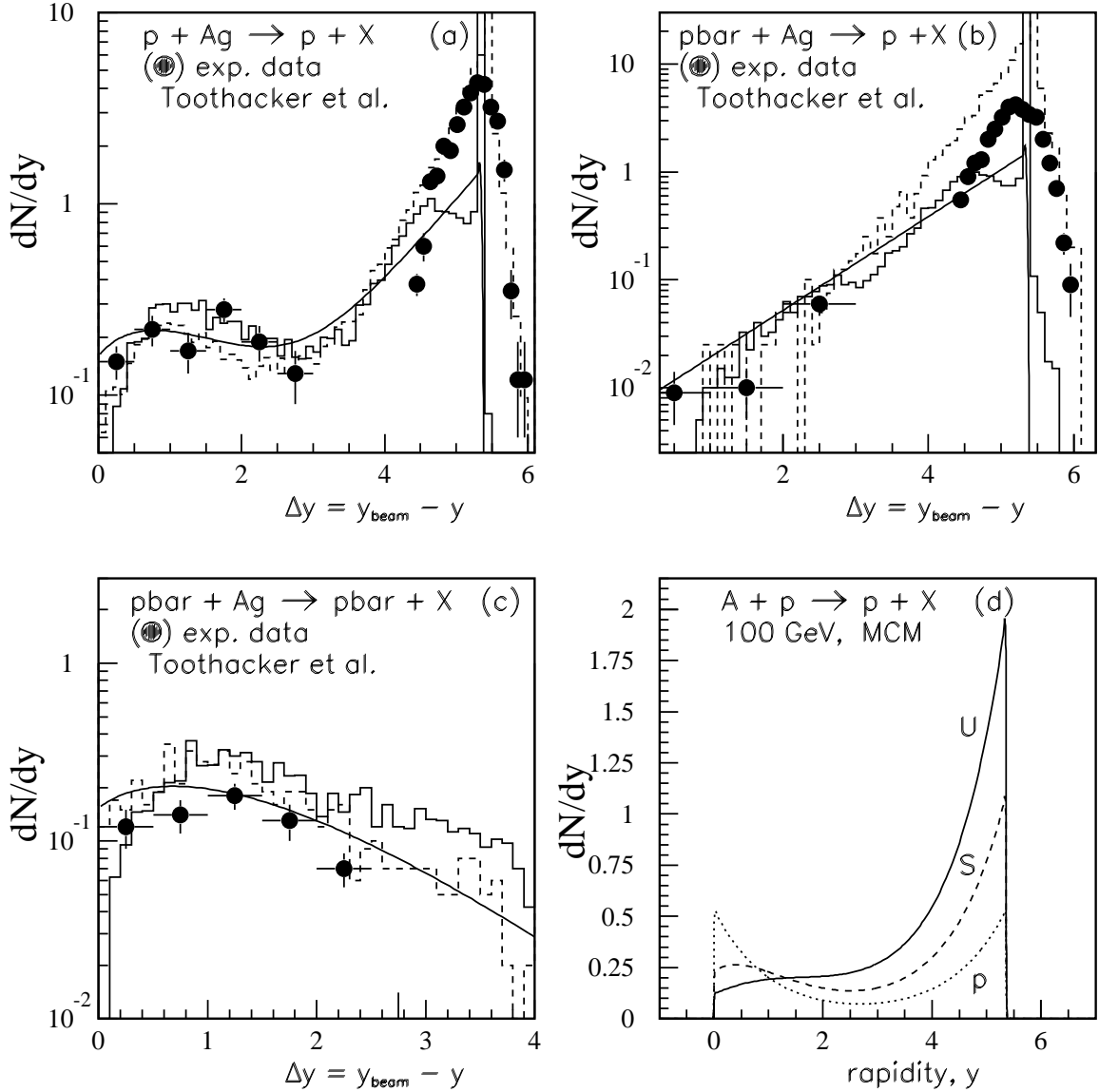}
\caption{Baryon transport in minimum bias 
 $ p,\bar{p} + Ag $ reactions at 100 GeV. 
The solid and dashed histograms display 
HIJING and VENUS model results respectively
as a function of
$\Delta y = y_{beam} - y $ with $ y_{beam} = 5.4 $.
 The smooth curve is
the result of the Multi-Chain Model.
The channel shown in part (b) 
isolates the target fragmentation region while (c) isolates the
projectile fragmentation region.
Part (d) shows the predicted $A$ dependence of the proton rapidity
distribution with $\alpha=3$ in the MCM.}
\end{figure}
\newpage
%Figure 3 
\begin{figure}[htb]
\vspace{1cm}
\hspace{0.5in}
\psfig{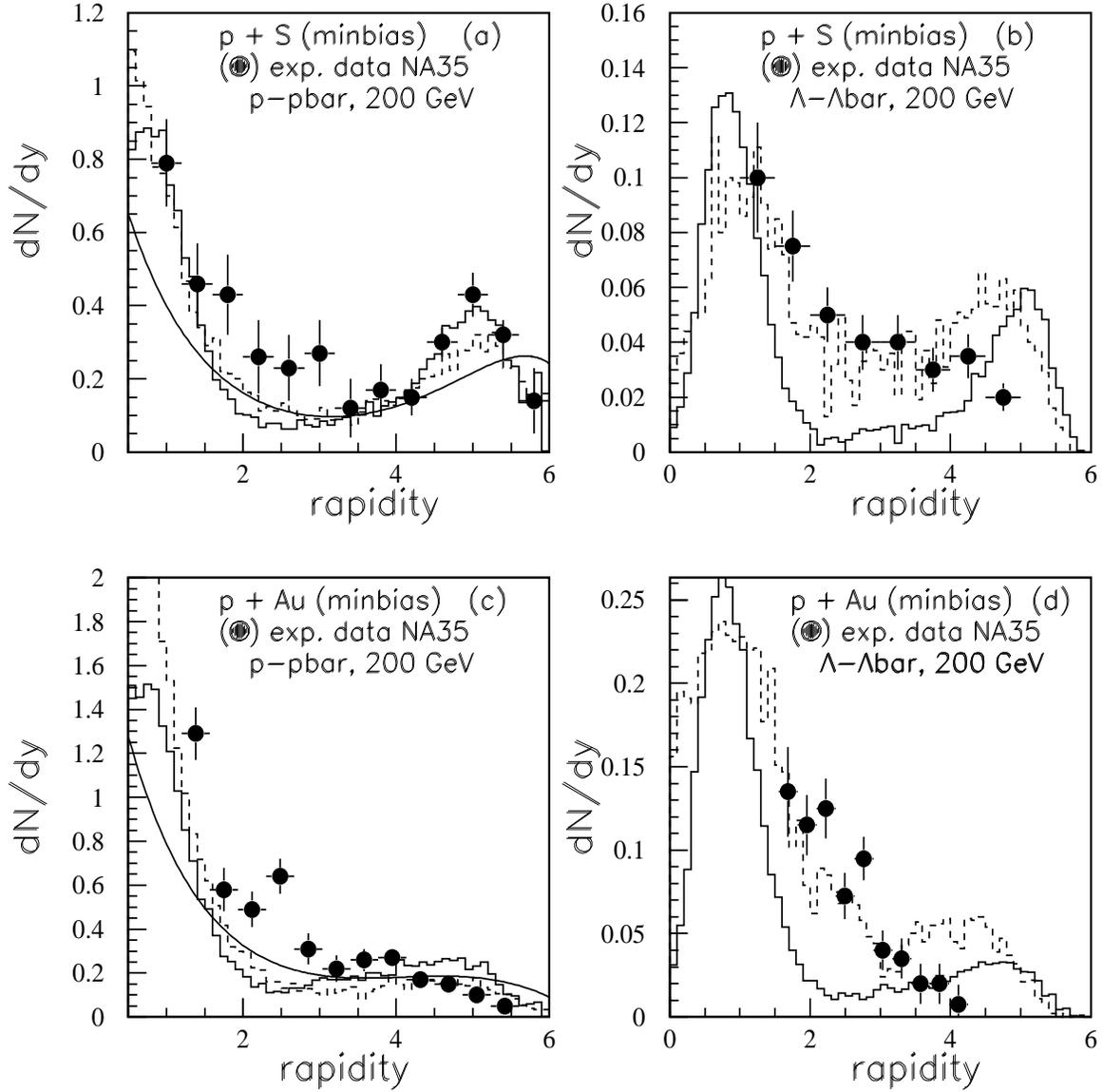}
\caption{Baryon number transport in minimum bias 
 $ p+ S,Au $ reactions\protect{\cite{na35_7}}  at 200 GeV. Curves are as
 in Fig.2. Here $y$ is the laboratory rapidity with $y>0.5$.
In parts (a) and (c) the $p-\bar{p}$ distributions are
shown while parts (b) and (d) correspond to the $\Lambda-\bar{\Lambda}$
distributions. Note that the data indicate significantly greater rapidity
shifts hyperons relative to non-strange-baryons in contrast to both VENUS and
HIJING calculations.}
\end{figure}

\newpage
%Figure 4 
\begin{figure}[htb]
\vspace{1cm}
\hspace{1in}
\psfig{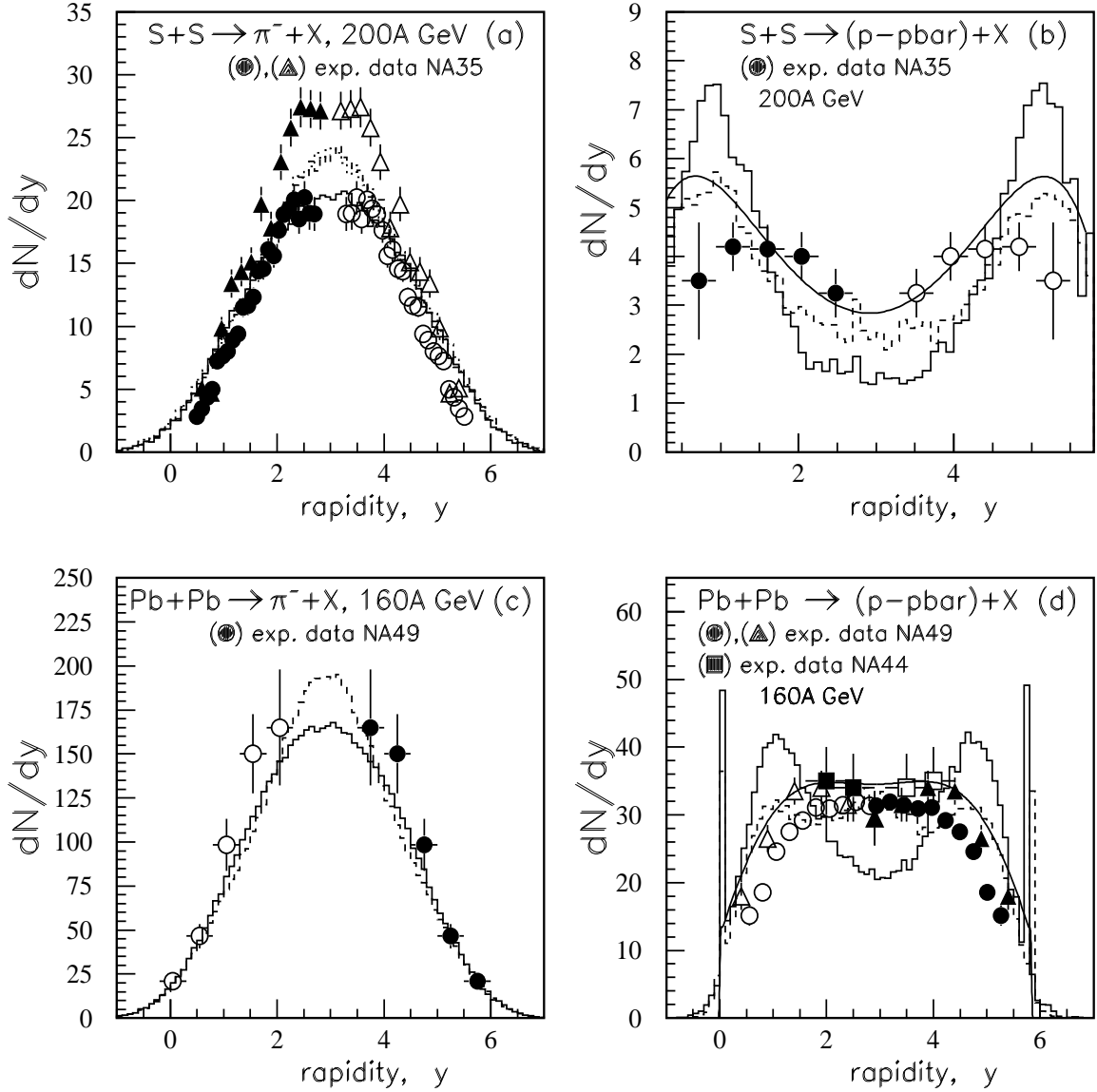}
\caption{Comparison of central $S+S$ at 200 AGeV (a,b) 
NA35\protect{\cite{na35_7,na35_4}} and central
$Pb+Pb$ at $160$ AGeV (c,d) (NA49\protect{\cite{na49_4,na49_1,na49_2}}
(solid triangles),
NA44\protect{\cite{xu_96}}(solid squares)) data with calculations.
Open circles and open squares are reflected data around mid-rapidity.
New preliminary analysis\protect{\cite{jacobs97}} of 
$p-\bar{p}$ are shown by solid dots in  part (d).
Solid and dashed histograms correspond to 
HIJING\protect{\cite{hijing}} and  VENUS\protect{\cite{venus}} models,
and continuous curves are predictions of the Multi-Chain Model
\protect{\cite{date}}.
}
\end{figure}

\newpage
%Figure 5 
\begin{figure}[thb]
\vspace{3cm}
\hspace{1in}
\psfig{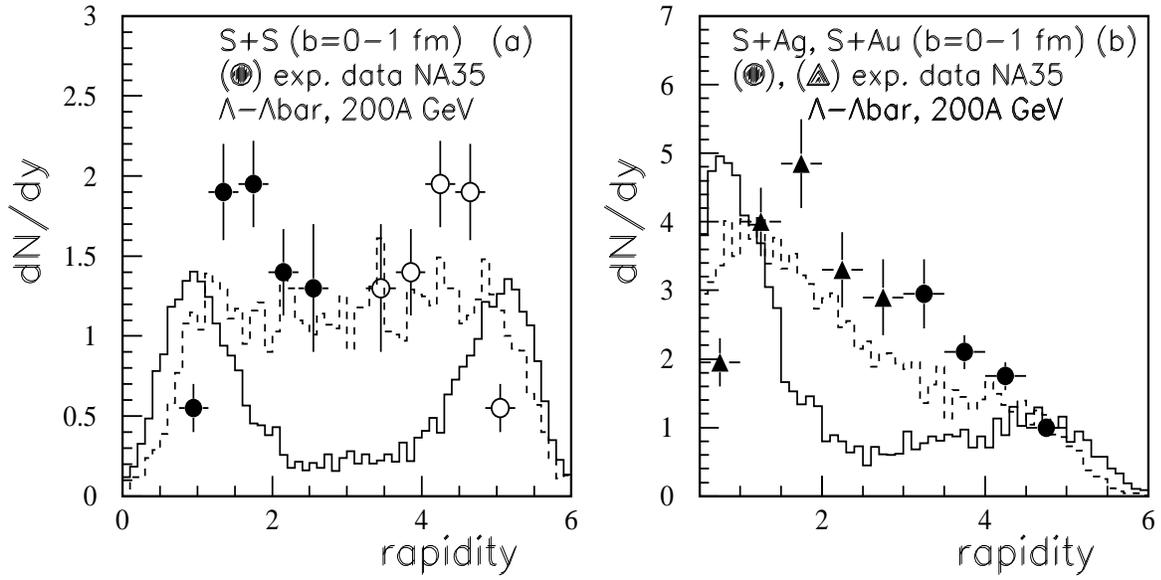}
\caption{Net Hyperon $\Lambda-\bar{\Lambda}$ rapidity 
distributions in central  $S+A$ at 200 AGeV (a,b) 
NA35\protect{\cite{na35_7}}.
Solid and dashed histograms correspond to 
HIJING\protect{\cite{hijing}} and  VENUS\protect{\cite{venus}} models.
}
\end{figure}

\newpage
%Figure 6
\begin{figure}[bht]
\vspace{1cm}
\hspace{1in}
\psfig{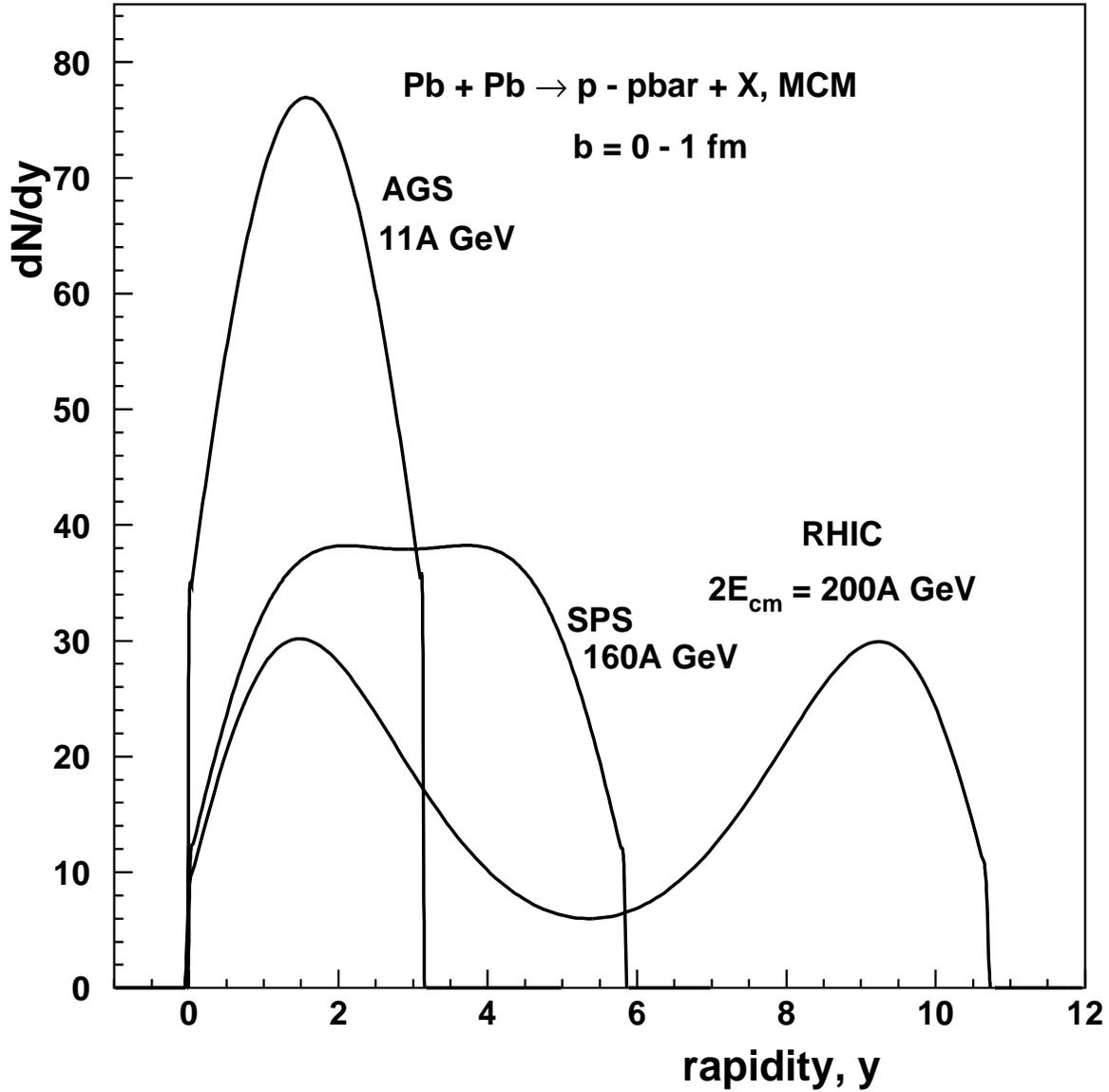}
\caption{The energy dependence of the net positive baryon number transport
in central $Pb+Pb$ reactions based on the MCM.
}
\end{figure}

\newpage
%Figure 7
\begin{figure}[htb]
\vspace{1cm}
\hspace{1in}
\psfig{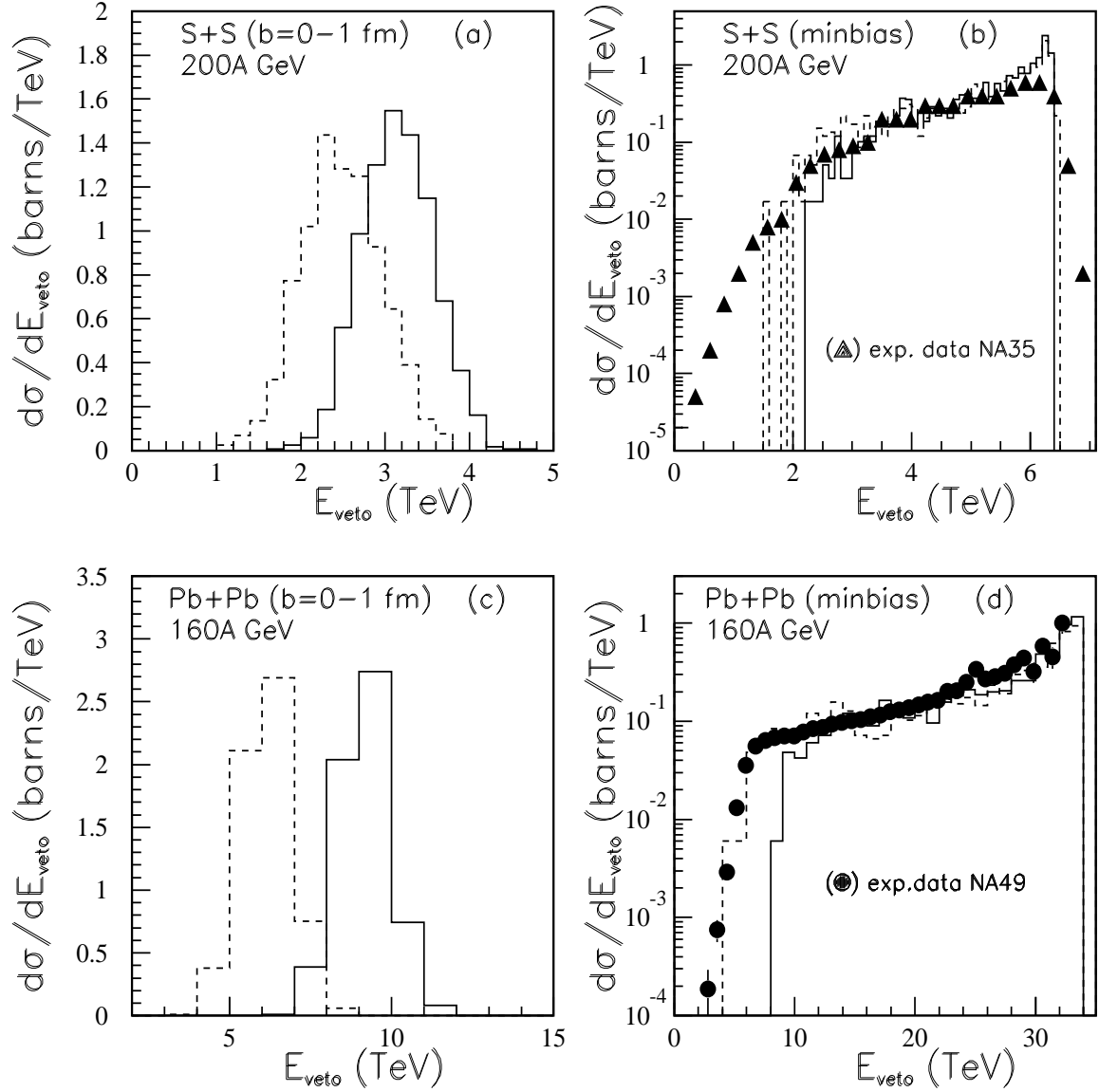}
\caption{The veto calorimeter cross section in 
$SS$\protect{\cite{na35_3}}
and  Pb-Pb\protect{\cite{na49_2}} collisions.
 Solid and dashed histograms refer
to HIJING and VENUS models.}
\end{figure}

\newpage
%Figure 8
\begin{figure}[htb]
\vspace{1cm}
\hspace{1in}
\psfig{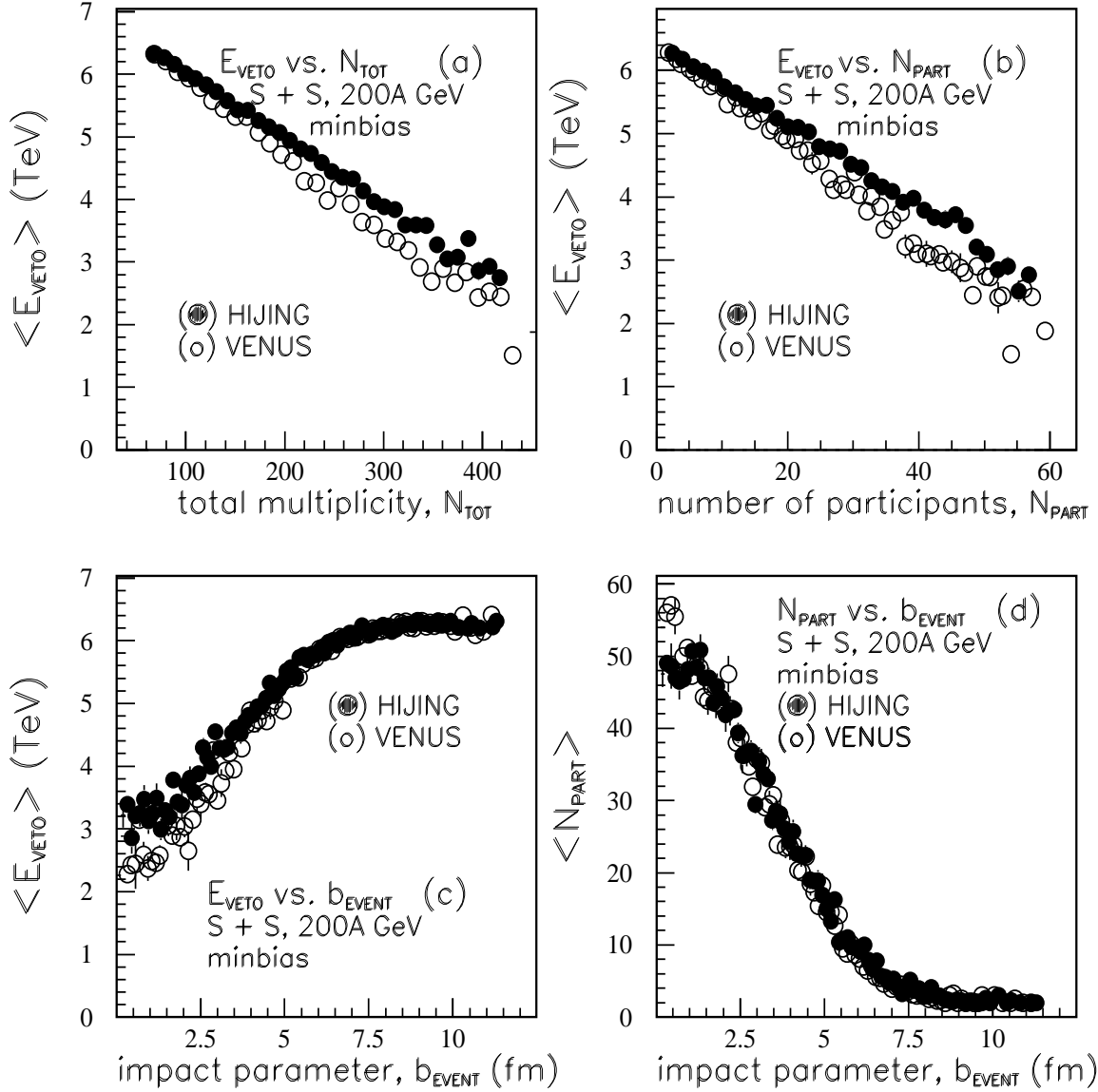}
\caption{Dependence of the  mean $E_{VETO}$ as a function
of (a) the total multiplicity, (b) number of participant (or wounded)
nucleons, and (c) impact parameter in $S+S$ reactions
comparing HIJING (solid) and VENUS (open) models.
Part (d) shows the mean participant number as a function of impact parameter.
}
\end{figure}

\newpage
%Figure 9
\begin{figure}[htb]
\vspace{1cm}
\hspace{1in}
\psfig{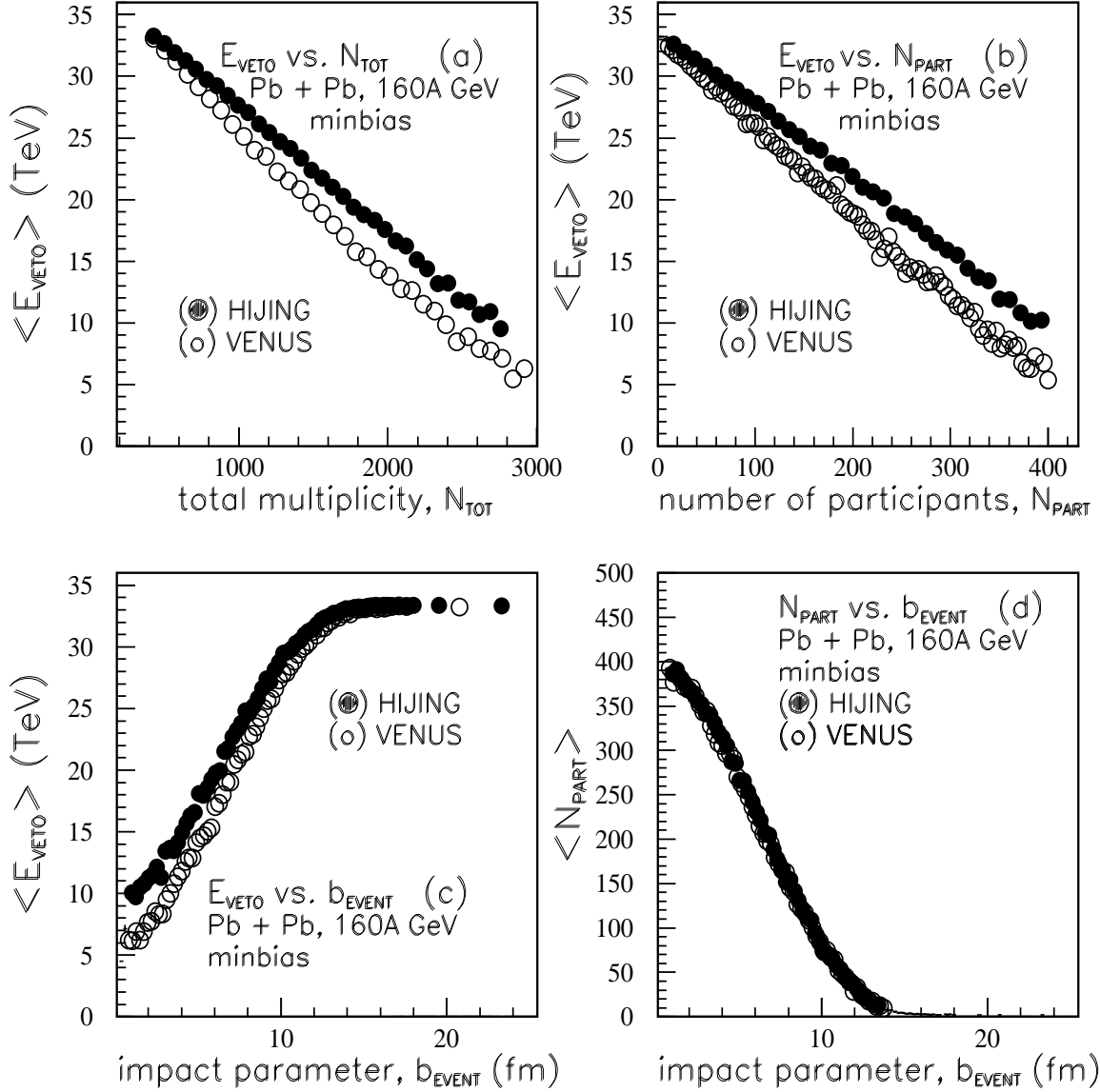}
\caption{Dependence of the  mean $E_{VETO}$ as a function
of (a) the total multiplicity, (b) number of participant (or wounded)
nucleons, and (c) impact parameter in $Pb+Pb$ reactions
comparing HIJING (solid) and VENUS (open) models.
Part (d) shows the mean participant number as a function of impact parameter.
}
\end{figure}

\newpage
%Figure 10
\begin{figure}[htb]
\vspace{1cm}
\hspace{1in}
\psfig{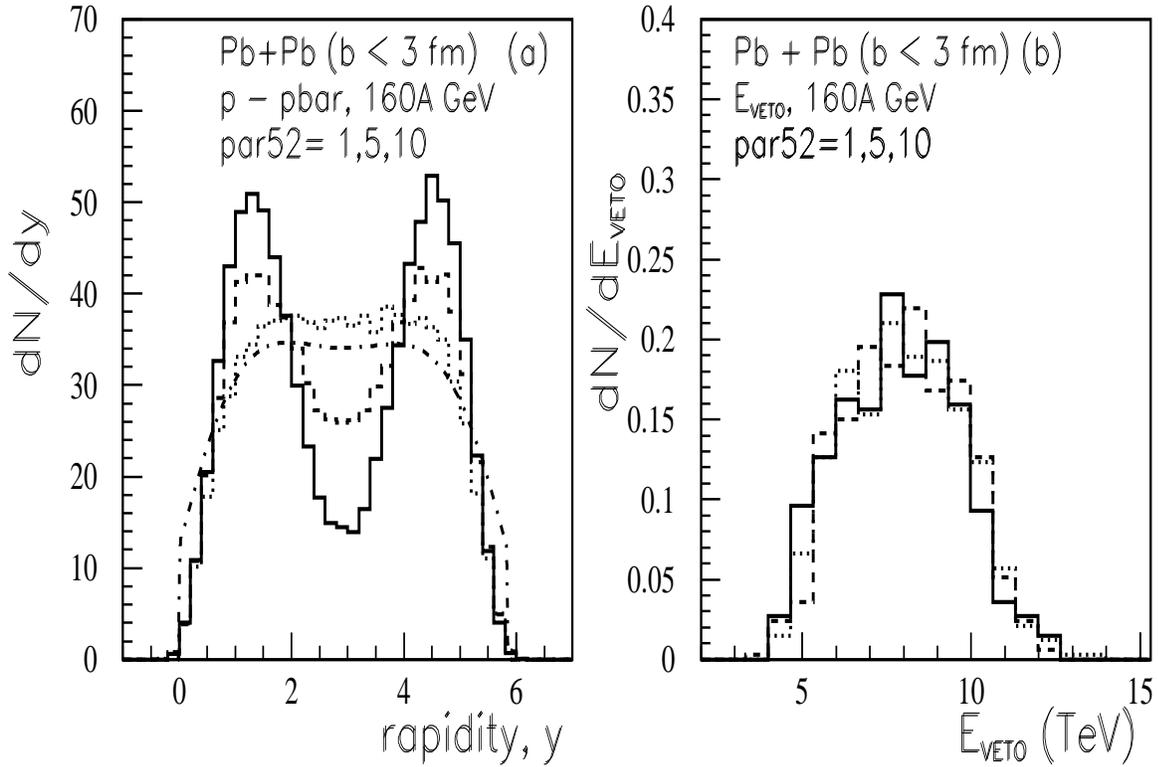}
\caption{(a) Variation of baryon transport using the 
ATTILA\protect{\cite{atti}} version of Fritiof\protect{\cite{fritiof}}.
The effect of changing the
parameter $p=par(52)$ in the jetset6.3\protect{\cite{jetset63}}
diquark fragmentation scheme is shown.
The default (solid) with $p=1$, and modified $p=5$ dashed
and $p=10$ doted histograms are shown compared to the MCM
prediction\protect{\cite{date}} as in Fig. 4d.
(b) The veto calorimeter cross sections and hence energy degradation
are uncorrelated to the baryon number transport in this model.
}
\end{figure}

\end{document}